# Sleep Disorder Diagnosis Using EEG Signals and LSTM Deep Learning Method

Mohammad Reza Yousefi1[1,*], Reza Rahimi[2],

[1]Department of Electrical Engineering, Na.C., Islamic Azad University, Najafabad, Iran
[2]Digital Processing and Machine Vision Research Center, Na.C., Islamic Azad University, Najafabad, Iran
* mr_yousefi@iau.ac.ir

**Abstract:** Diagnosing brain disorders, particularly sleep disorders, has emerged as a critical area of research in both basic sciences and engineering. Sleep disorders encompass a range of conditions, including excessive or insufficient sleep, frequent awakenings, difficulty falling asleep, and challenges in reaching deep sleep stages. A key application of sleep disorder diagnosis lies in supporting the medical community through the development of advanced algorithms and models capable of accurately identifying these disorders based on individual brain signals, such as those captured by electroencephalography (EEG). These sophisticated approaches facilitate the creation of personalized treatment programs, paving the way for more targeted and effective interventions. Historically, statistical pattern recognition methods were regarded as the gold standard for analyzing brain signals and identifying sleep disorders. However, with the rapid evolution of deep learning, these traditional methods have been increasingly supplanted by more advanced deep learning techniques. To fully harness the potential of these emerging technologies, further research is essential to explore the application of deep learning in analyzing brain signals for sleep disorder identification. Such efforts can address existing knowledge gaps and expand our understanding in this field. In this study, a comprehensive public database was utilized, consisting of 197 full-night sleep recordings from participants aged 25 to 101 years, including both male and female subjects. Following rigorous preprocessing stages—such as noise elimination and signal refinement—critical features were extracted and analyzed. Using deep learning techniques, particularly Long Short-Term Memory (LSTM) neural networks, the researchers achieved a remarkable accuracy rate of 93.3% in distinguishing between healthy and diseased classes. By incorporating advanced fusion techniques, the classification accuracy was further improved to 95%. Notably, the computational efficiency of these deep learning networks makes them highly promising for clinical applications in sleep disorder diagnosis. Their relatively low processing time suggests that such algorithms could be invaluable in clinical settings, enabling rapid and precise detection of sleep disorders.

## 1. Introduction

Sleep has always been a subject that has captivated the minds of researchers and scientists. Research shows that sleep evolved for better efficiency in the evolutionary process, and human sleep environments have undergone significant changes over time. In the modern world, with technological advancements and lifestyle changes, Sleep Disorders (SD) have increased. Recent research in sleep physiology and pathology, particularly regarding its effects on human health and behavior, has experienced remarkable growth. Although historical information about sleep patterns and disorders in prehistoric times is limited, studying sleep patterns in animals and other evidence can lead to some insights [1]. The medical history of sleep in the United States began with the establishment of the Sleep Psychophysiological Studies Association in 1961 and continued with the creation of clinical sleep centers in the 1970s and the establishment of accreditation processes by the Sleep Disorders Association [2]. According to a 1979 medical institute report, approximately 50 million adult Americans (about one-third of the population) suffer from sleep problems, with nearly 10 million seeking medical attention annually, and half of them receiving sleep medications [2]. The human body has evolved to synchronize its physiological processes with the natural light-dark cycle, enabling proactive responses to sleep and wake periods. Disruption of this circadian rhythm can lead to significant functional disorders [3]. Moreover, studies on mice demonstrate that long-term sleeplessness can be fatal within a month. Historically, sleeplessness has been used as a method of torture, highlighting its profound impact on physical and mental health [4]. These findings emphasize the vital importance of sleep for human well-being.

Sleep is a complex neurological state whose primary purpose is to provide rest and restore body energy [5]. Humans spend approximately one-third of their lives sleeping, and this state offers specific benefits such as memory enhancement, energy restoration, and support for learning processes [6]. Human sleep is divided into two types: REM and NREM, with NREM sleep comprising four distinct stages [7]. In 2007, the American Academy of Sleep Medicine revised the sleep staging system, merging stages S3 and S4 into N3 [8]. Sleep stages include N1 (light sleep lasting 5-10 minutes), N2 (50% of nightly sleep), N3 (deep, restorative sleep), and REM (dreaming phase). Each sleep cycle lasts 90-120 minutes and repeats 4-6 times per night. Today, sleep disorders are recognized as significant human problems. Consequently, the National Sleep Foundation (NSF) in Virginia, USA, suggests that determining an individual's daily sleep requirements depends on their lifestyle. However, based on their research, they have generally established appropriate sleep durations for healthy individuals.

Table (1) suitable sleep time for healthy people by each group [9]

| Older adults | Youth and adults | Teenagers | Primary school children | Preschool children | Young children | Infants | Babies |
|---|---|---|---|---|---|---|---|
| 7-8 Hours | 7-9 H | 8-10 H | 9-11 H | 10-13 H | 11-14 H | 12-15 H | 14-17 H |

Sleep duration varies individually. Sleeping outside normal ranges may indicate health issues or deliberate sleep restriction. Recommendations are based on scientific evidence for healthy people without sleep disorders.

Sleep disorders are recognized as conditions associated with changes in normal sleep patterns or behaviors. These disorders can lead to problems in daily functioning and cause discomfort [10]. According to reports, approximately 40 million people in the United States suffer from chronic sleep and wakefulness disorders. The most common sleep disorders include sleep apnea, insomnia, narcolepsy (sleep attacks), parasomnias, and Restless Leg Syndrome (RLS). Sleep disorders annually result in 38,000 cardiovascular deaths. These disorders impact driving, social activities, and individuals' work [11]. Sleep deficiency or disruption has irreparable consequences on life quality, occupational-educational performance, immune system, and mental health. These consequences include memory disruption, increased accident risks, and weakened body immune systems. Individuals with less than standard sleep are at high risk of heart diseases, diabetes, and obesity. Sleep disorders encompass a wide range of problems affecting individuals' life quality. Insomnia, the most common sleep disorder, is characterized by difficulty falling asleep, maintaining sleep, or low sleep quality, potentially arising from stress, anxiety, or psychological issues. Restless Leg Syndrome is a neuro-sensory disorder accompanied by unpleasant sensation and strong urge to move legs during rest, typically intensifying at night. Obstructive sleep apnea is a serious disorder characterized by intermittent breathing stops during sleep, potentially leading to reduced sleep quality, daily fatigue, and cardiovascular problems. Parasomnias are a group of behavioral disorders occurring during sleep, including sleepwalking, sleep talking, nightmares, and motor disturbances that can disrupt the individual's and family members' sleep. According to the World Health Organization, the prevalence of sleep disorders is generally increasing worldwide, especially in Asia and Africa. Factors such as daily stress, lifestyle quality and changes, excessive technology use, and increasing psychological problems have contributed to this increase [12].

The human brain, with approximately 100 billion neurons, is one of the most complex biological structures where neurons communicate through synapses. This intricate network is responsible for information processing, movement control, emotions, and decision-making. Neurons are divided into three main categories: sensory neurons (receiving information from the environment), motor neurons (sending messages to muscles), and interneurons (processing information within the central nervous system).

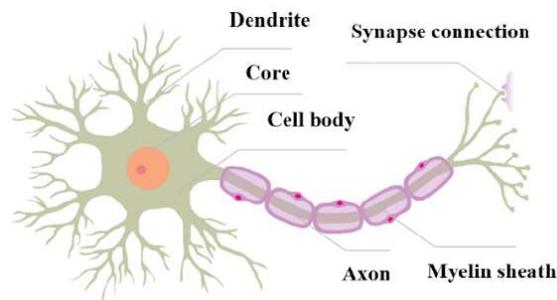

Figure (1) structure of nerve cell [13]

Studies show that sleep occurs at the level of neuronal groups and promotes synaptic structure preservation. Sleep deprivation causes learning problems and sleep pattern disruption by altering synaptic plasticity, increasing orexin neuron activity, and affecting circadian rhythm gene expression [14]. According to research, sleep deprivation and sleep disorders in humans can have significant effects at the neuronal level of the brain. One of the main mechanisms through which sleep deprivation can affect the brain is by altering synaptic plasticity. Synaptic plasticity refers to the ability of synapses to strengthen or weaken over time and plays a vital role in learning and memory. When sleep patterns are disrupted, these plasticity processes may be disturbed, leading to learning problems [15]. These disturbances also increase the activity of orexin neurons, which play an important role in regulating the sleep-wake cycle. Sleep deprivation can cause increased activity of these neurons, which can lead to disruption of normal sleep patterns. This can cause insomnia, daytime sleepiness, and disruption of the normal human sleep pattern [16]. Additionally, sleep disorders can affect the expression of key genes in neurons, especially those associated with circadian rhythms. Circadian rhythms regulate natural sleep-wake cycles and are essential for optimal brain function. When the expression of these genes changes due to sleep disorders, it can lead to disruption of sleep-wake cycles, which in turn can affect a wide range of brain functions [17]. In sleep research, especially in laboratory settings such as sleep clinics, the technique of polysomnography (PSG) is used to measure sleep [18]. Polysomnography (PSG) technique is used to diagnose sleep disorders by recording brain activity, muscle activity, eye movements, and other physiological indicators during sleep. Sufficient and quality sleep is essential for brain health, and sleep deprivation has significant negative consequences.

The electroencephalogram (EEG) signal is a complex and widely used signal for recording the electrical activity of the brain, also known as the brain signal [19–36]. These signals result from the activity of neurons in the brain, which are non-invasively recorded by electrodes on the scalp surface and can provide important information about brain function, sleep, concentration, and even neurological disorders such as epilepsy or sleep disorders [37]. EEG signal measurement utilizes surface electrodes placed on the scalp following the International 10-20 system, a globally standardized electrode placement method. This system determines precise electrode locations by dividing distances between key anatomical landmarks (nasion, inion, and preauricular points) into 10% and 20% intervals. The electrodes are typically made of Ag/AgCl (silver-silver chloride) due to their electrochemical stability and low noise characteristics. Conductive gel is applied to enhance electrical contact between electrodes and scalp. Since recorded signals are extremely weak (microvolt range), high-gain amplifiers are required, along with analog and digital filters to eliminate environmental noise and improve signal quality [38].

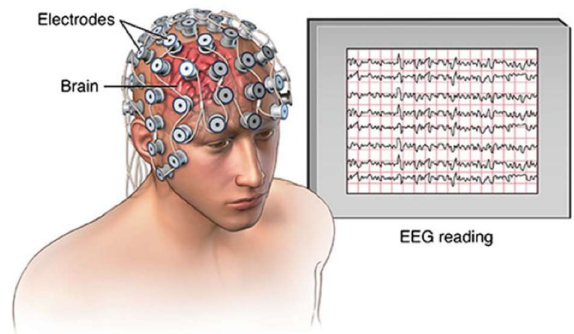

Figure (2) of the recording of the electroencephalogram signal [39]

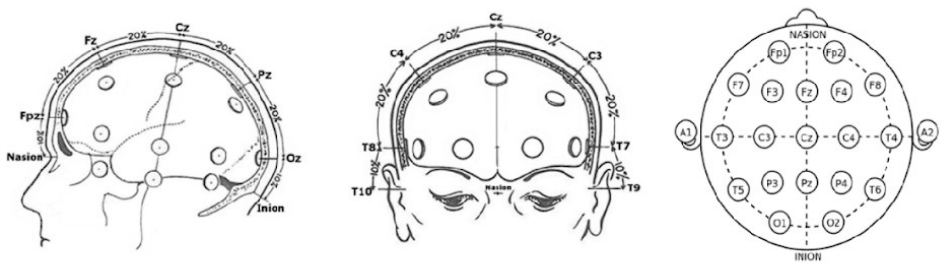

Figure (3) showing the 10-20 electrode system from different angles [40,41]

Invasive methods of EEG signal recording, such as using implantable intracranial electrodes, although they provide higher accuracy in recording electrical activities of the brain, are accompanied by challenges. These methods require surgery to implant electrodes in the brain, which carry risks such as infection, bleeding, and tissue damage [42]. Additionally, past invasive methods faced limitations including a small number of electrodes, limited coverage of brain regions, and difficulty in interpreting complex data. Other problems included immune reactions to foreign materials, gradual degradation of electrodes, and the need for repeated surgeries to replace or adjust them [43]. These limitations not only restricted the widespread use of these methods but also created significant ethical and practical challenges. However, recent advances in material technology and electrode design have reduced some of these problems and provided the possibility for longer-term and higher-quality recording. The use of EEG signals in disease diagnosis, especially sleep disorders, has gained increasing importance. This non-invasive method enables the recording of electrical activities of the brain and provides valuable information about brain function in various states, including sleep. Analysis of EEG patterns can help identify neurological abnormalities and sleep disorders such as sleep apnea, insomnia, and narcolepsy. This method allows physicians to provide more accurate diagnoses and prescribe more effective treatments for patients [44]. In addition to sleep disorders, the use of EEG signals in diagnosing a wide range of neurological and psychiatric diseases is also of special importance. This method has found widespread application in the diagnosis and management of diseases such as epilepsy, Alzheimer's, Parkinson's, and autism spectrum disorders [45]. Therefore, EEG signals provide useful information about brain activities, and the characteristics of EEG signals can act as an efficient tool for diagnosing and predicting neurological disorders. EEG signals record the electrical potentials of the brain that are characterized by different frequencies [46]. Although electroencephalogram signals have many applications in medical centers from a signal recording perspective, their processing algorithms in the field of feature extraction and pattern extraction are in their early stages and need further research. In particular, diagnostic support systems that can diagnose neurological diseases from EEG signals have received less attention from researchers, and no precise preprocessing technique has been developed for this signal regarding what frequency range it operates in and how it works [39]. EEG signal recordings are often contaminated by artifacts such as eye movements and blinking, which cause electrical potentials in the eye region. These artifacts can affect the EEG signal and make data interpretation difficult [47]. For this reason, we need an algorithm that, after preprocessing the signals, can process the signals of healthy and unhealthy individuals with high accuracy and in the shortest possible time and identify the person with sleep disorder. Also, this algorithm should be such that it is not dependent on signal preprocessing and can provide appropriate performance if contaminated with artifacts and noise. Signal processing methods are categorized into two groups: deep learning-based research and classical methods (machine learning). In classical methods, the first stage is the use of vital data. In this type of signal processing, features are manually selected, and with their selection, error increases significantly; it is also worth noting that this method is very sensitive to noise. The absence of noise and artifacts in a signal is impossible, and for this reason, we need methods that are not sensitive to noise. Therefore, we will use deep learning networks that perform all operations related to signal processing in the shortest time and automatically, which is also a very high advantage for clinical applications.

**2. Deep learning networks**

The human brain is the inspiration for the structure of deep neural networks. As mentioned at the beginning of this chapter, the brain is composed of fundamental units called neurons, whose functions can be understood by examining their main components. Each neuron consists of several parts: First, dendrites, which are branches that receive neural signals carrying information. The thickness of dendrites indicates the importance of the received information; thicker dendrites transmit more important information. The second part is the cell nucleus, which processes the information received from dendrites. This processed information is transmitted through the axon, which is the third part of the neuron. The axon is the pathway for transmitting information from the nucleus to the final part of the neuron, the synapse. Finally, the synapse is where processed information exits the neuron and is transferred to other neurons. The artificial neuron model is designed by patterning after the function of natural or biological neurons. In biological neurons, input signals are received through dendrites and collected and processed in the cell nucleus. The result of this processing is transmitted to other neurons through the axon and synapses. The goal of developing artificial neurons is to create an algorithm with efficiency at least equal to, and preferably beyond, the performance of the human brain [48]. Figure 4 shows the architecture of artificial and natural models.

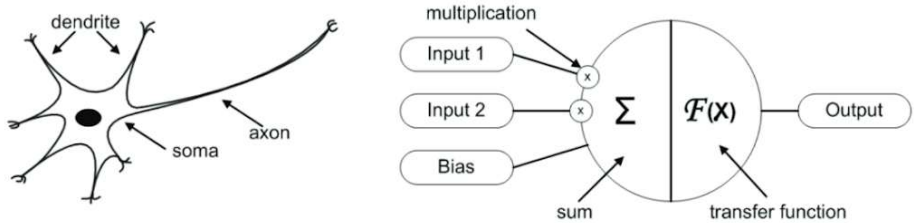

Figure (4) design of biological neuron and artificial neuron [49]

Figure (4) presents a comparison between the structure of natural and artificial neurons. As mentioned, in natural neurons, dendrites are responsible for receiving input information. In the artificial model, inputs 1 to n perform the role of dendrites. The similarity between dendrites and artificial inputs lies in receiving input information. In natural neurons, dendrite thickness indicates information importance, which is simulated in artificial neurons using weights x1 to xm. In biological neurons, information is processed in the cell nucleus and transmitted through axons, while artificial neurons sum inputs in a weighted manner. Artificial neural networks are machine learning algorithms that discover hidden patterns in data. Deep learning has abandoned traditional feature extraction methods and instead of manual data preparation, algorithms train directly on raw data. This has transferred the responsibility of feature extraction from humans to computer algorithms. Deep artificial neural networks with more than two layers are considered an advanced generation of traditional networks.This complex architecture enables hierarchical feature extraction; meaning complex features are extracted from combinations of simpler features. The high efficiency of this method when dealing with large volumes of training data has made it a powerful tool in areas such as sleep studies, with massive data [50]. Although the concept of deep learning has been around since the 1980s, recent advances in two areas have expanded its application:

1) The ability of this method to process and classify massive labeled datasets
2) Access to high computational power, especially parallel graphics processors (GPUs) that address the heavy computational requirements of deep learning.

Next, two widely used models of neural networks based on deep learning will be examined.

**2-1- Recurrent Neural Network**

Recurrent Neural Networks (RNNs) are artificial neural networks designed for processing sequential data like text, audio, and time-series. These networks use an internal memory mechanism that considers past information when processing current inputs. Unlike regular networks, RNNs have recurrent connections that enable using information from previous steps. In a simple RNN, each hidden layer node receives input from both the previous layer and the output of the corresponding node from the previous time step.

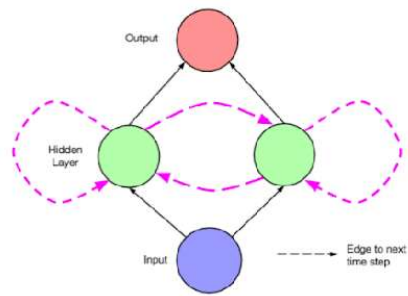

Figure (5) Recurrent neural network model [51]

Figure (5) shows a simple RNN where activation transfers through solid lines like feedforward networks at each time step t. Dashed lines connect nodes from time t to t+1. The network includes input, hidden, and output layers. Besides inter-layer activation transfer, activation also moves from hidden layer at time t to t+1, demonstrating the recurrent nature. This structure enables information retention over time. Simple RNNs have limitations including vanishing gradient problems that hinder long-term temporal dependency learning. The main advantage of RNNs is effective modeling of long-term relationships in time-series data, unlike traditional Markov models. RNNs overcome these limitations using recurrent connections and advanced architectures like LSTM.

**2-2- Long Short-Term Memory Neural Network**

As mentioned, the LSTM neural network is an advanced version of the RNN neural network, and today this neural network has found extensive applications [52]. This network has shown remarkable performance not only in text and speech processing but also in image processing and especially signal processing. LSTM networks demonstrate remarkable versatility, potentially surpassing CNN applications in certain image processing scenarios, despite CNNs being the predominant choice for visual tasks like face recognition. LSTM stands out as a sophisticated artificial intelligence system capable of learning and comprehending intricate patterns. With recent deep learning breakthroughs, LSTM has emerged as a fundamental tool in AI development. Researchers actively explore LSTM's potential to deliver innovative solutions across diverse data processing challenges. As a specialized RNN variant, LSTM is specifically engineered to capture and model long-term dependencies within time series data, making it particularly valuable for sequential pattern recognition tasks. This neural network was first introduced by Hochreiter and Schmidhuber in 1997 [53] and has since become one of the most popular neural network architectures for processing sequential (time series) data. An LSTM cell consists of three main gates: the forget gate, the input gate, and the output gate, which are shown in the block diagram of the LSTM neural network in Figure (6).

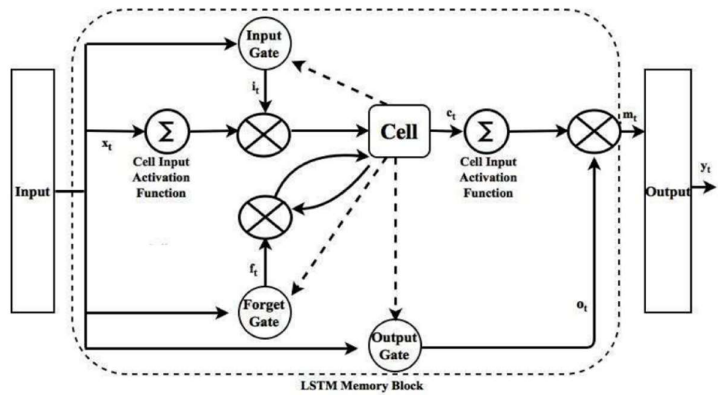

Figure (6): LSTM network block diagram with three main gates: input gate, forget gate and output gate [53]

These gates operate as follows:
1) **Input Gate:** Determines what new information should be stored in cell memory and acts as a filter that preserves relevant information.
2) **Forget Gate:** Decides which information from the previous cell state should be forgotten, allowing the network to clear unnecessary information and free up space for newer data.
3) **Output Gate:** Controls what information from the current cell state should be transferred as output to the current hidden state.

All three gates combine current input information and previous hidden state to produce values between 0 and 1.

**3- Proposed method**

In this method, we try to detect sleep disorders through EEG signals using Long Short-Term Memory neural networks. Figure 7-1 shows the block diagram of this method, and we will examine each component in detail in the following sections.

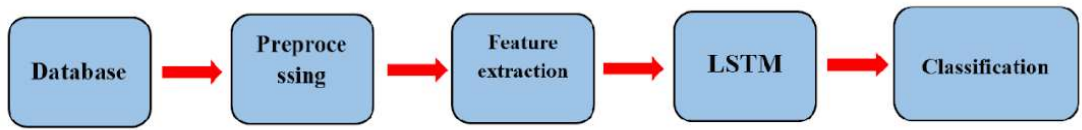

Figure (7) block diagram of the proposed method

### 3-1- Database

The Sleep EDF database is one of the reputable collections in the field of sleep analysis that has been specifically designed for research in sleep science and biomedical signal processing. This database is publicly accessible on the PhysioNet website. The database consists of 197 full-night polysomnography recordings from participants, including 153 recordings from healthy individuals who have no diseases or medication use, and 44 recordings from patients who have mild difficulty falling asleep. The age range of participants in this database is 25-101 years, and the sampling frequency of the recordings is 100 Hz. Additionally, the corresponding hypnograms (sleep patterns) have been manually scored by trained technicians [54]. The PSG recordings of healthy individuals are 7-channel, containing two EEG channels (Fpz-Cz channel and Pz-Oz channel), one horizontal EOG channel, one chin EMG channel, one respiratory channel, body temperature, and an event marker. The recordings of patients are 5-channel, including two EEG channels (Fpz-Cz channel and Pz-Oz channel), one horizontal EOG channel, one chin EMG channel, and an event marker. In the first stage of implementing our proposed method, we retained the two EEG channels and removed the remaining additional channels from the recordings. Figure 8 shows the two recorded EEG channels from one of the patients.

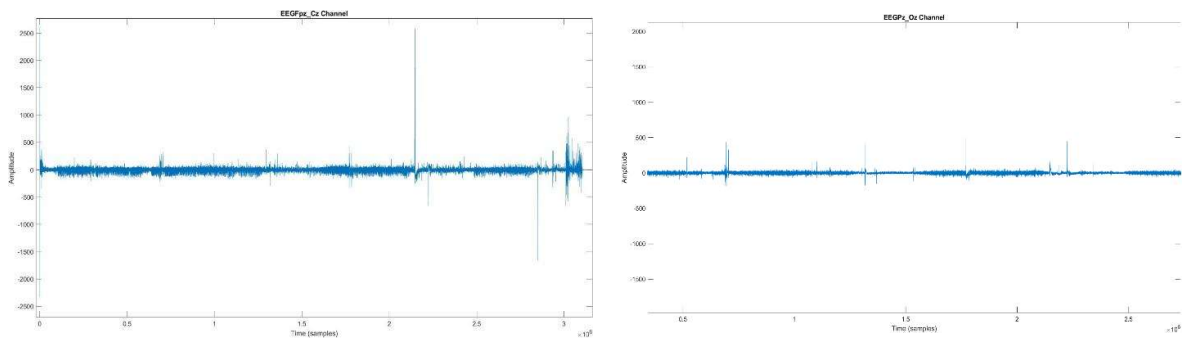

Figure (8) Pz-Oz channel of the EEG signal related to one of the sick people and Fpz-Cz channel of the EEG signal related to one of the sick people

### 3-2- Preprocessing

EEG signals are widely used to examine electrical activities of the brain, especially in the field of sleep disorders. However, these signals are affected by various noises and artifacts that can make data analysis and interpretation difficult and problematic. Noises are usually divided into two categories: internal noises and external noises. Internal noises include unrelated brain activities, such as eye artifacts, muscle artifacts, and changes caused by breathing. While external noises are usually caused by environmental factors such as power line noise. The noise removal operation from the data is performed in two stages:

First stage: Given that power line noise is observed in the frequency range of 50 Hz, the presence of this noise can have a significant impact on the EEG signal and final accuracy. However, it should be noted that neural networks are more resistant to noise compared to statistical methods, but removing noise from the signal increases accuracy and makes the simulation more precise. To remove this noise, one of the common methods is using digital filters, and in this work, a notch filter has been used to remove power line noise in the 50 Hz range. Notch filters are specifically designed for frequencies in specific ranges; these filters are able to reduce the effect of power line noise in the signal by removing a band of frequencies (for example, 48 to 52 Hz).

Second stage: Applying a 5th-order Butterworth bandpass filter in the range of the signal used in this work. The Butterworth bandpass filter is one of the types of digital filters known for its flat and non-oscillating response in the passband. This filter can be used to remove unwanted frequencies and artifacts in the EEG signal.

### 3-4- Feature Extraction

In the analysis of EEG signals related to sleep disorders, one of the key stages is windowing and feature extraction. These steps help extract useful information from complex and noisy signals. Windowing refers to the process of dividing EEG signals into smaller sections (windows). This is done for two main reasons: Local analysis: EEG signals typically have rapid and non-linear changes. By dividing the signal into smaller windows, local changes can be better identified and different sleep patterns can be analyzed. And data volume reduction: Sleep-related EEG signals can be very large, and processing them as a whole may be time-consuming and impractical. With windowing, the volume of data can be reduced, and analysis can be performed on each window separately. The length of each window is typically chosen between 1 to 30 seconds, and in this proposed method, the length of each window is considered to be 10 seconds. After windowing, the next stage is feature extraction. In this stage, important features related to sleep disorders are extracted from each window and substituted. These features include frequency features, statistical features (such as mean, standard deviation, etc.), and non-linear features (such as entropy, etc.).

### 3-5- The Proposed LSTM Network

In this proposed method, Long Short-Term Memory (LSTM) neural network has been used. LSTM is a very powerful neural network that is mostly used for time series and sequential data, and since sleep disorder data is also considered a type of sequential data, we have used this network. After preprocessing and feature extraction stages, the data was divided in a 70-30 ratio, with 70 percent selected for training and 30 percent for testing. In this proposed network, 12 LSTM layers (LSTM blocks) have been used, which allows the network to extract more complex features from the data. Each block can help learn one level of features, and by increasing the number of blocks, the model's ability to learn complex patterns increases. The number of hidden layers in this network was set to 200, with 200 neurons placed in each hidden layer, giving the network high power in learning and feature extraction. Additionally, the learning rate was chosen as 0.01. This value allows the model to move slowly and more accurately toward the minimum of the cost function. An excessively high learning rate may lead to oscillations and non-convergence, while a very low learning rate causes the model to take a long time for training. The Dropout Layer value in this method is 0.2, which is used to prevent overfitting in the model. By randomly removing 20 percent of neurons in each layer during training, the network is able to learn more general features and prevent excessive dependence on training data. The value of 125 was chosen for the number of iterations, meaning that during the training process, the model will run 125 times on the training data. Selecting an appropriate number of iterations is important; too few may lead to insufficient learning, and too many may lead to overlearning. Finally, the SGDM optimizer was used, which has been developed as an advanced method of the basic Stochastic Gradient Descent (SGD) algorithm and is designed to improve convergence speed and reduce fluctuations in the training process.

**3-6- Fusion Technique**

The fusion technique refers to combining the results of multiple models or predictions to improve the accuracy and efficiency of a machine learning system. In our proposed method, we specifically used this technique for the LSTM neural network to increase the model's ability to classify healthy and sick individuals. In the stages of using the fusion technique in the LSTM network, first, the LSTM model is trained on EEG data. This model, due to its unique features, including memory cells and gate mechanisms, is able to identify long-term dependencies in time-series data and is therefore very effective for analyzing EEG signals. After completing the training process, the LSTM model is applied to test data to predict patients' sleep status. In the next stage, to increase the accuracy of predictions, results from several different runs of the LSTM model (with changes in parameters or network structure) are combined. This combination can be done through methods such as averaging or voting, where the strengths of each model are aggregated into a final prediction. Finally, after applying the fusion technique and majority voting method, the overall accuracy of the system is evaluated, and the results show a significant increase in accuracy in diagnosing sleep disorders.

**4- Results**

The validation of this proposed method for the used classifier was compared using the criteria of accuracy, sensitivity, and specificity obtained from the confusion matrix, and finally the optimal structure is introduced.
The parameter TP (True Positive) indicates the number of samples that have been correctly identified as positive. In other words, TP represents cases that the model has correctly predicted to belong to the positive target class. The parameter TN (True Negative) is the number of samples that have been correctly identified as negative, meaning the model has correctly predicted these cases belong to the negative class. The parameter FP (False Positive) indicates the number of samples that have been incorrectly identified as positive, in other words, the number of cases the model predicted to be positive but are actually negative. And the parameter FN (False Negative) is the number of samples that have been incorrectly identified as negative, meaning the model predicted these cases to be negative but they are actually positive. In fact, the closer the values of these parameters are to one, the closer the results are to the ideal answer. The simulation was performed using Matlab R2022b software with an Nvidia GTX-960M graphics card and 32 GB of RAM.
Figure 9 shows the confusion matrix resulting from the proposed method for sleep disorder detection, which indicates that the proposed neural network has been able to classify the two classes of healthy individuals and patients with an accuracy of 93.3 percent.
Figure 10 shows the confusion matrix resulting from the proposed method after using the fusion technique for sleep disorder detection, which indicates that the used technique has been able to classify the two classes of healthy individuals and patients with an accuracy of 95 percent.

Figure (9) Confusion matrix resulting from the proposed method with LSTM neural network

Figure (10) confusion matrix resulting from the proposed method after using the fusion technique

## 5- Discussion and Conclusion

Sleep is one of the most fundamental physiological needs of humans that plays a vital role in maintaining physical and mental health. Accordingly, humans spend approximately one-third of their lives sleeping, and scientific studies indicate that adults need 7 to 9 hours of daily sleep. Sleep deficiency or disorders can have irreversible consequences on quality of life, occupational-educational problems, immune system, and mental health, leading to memory impairment, increased risk of accidents, and weakened immune system. People who sleep less than 6 hours per night are at higher risk for heart disease, diabetes, and obesity. Additionally, sleep disorders can cause neuropsychiatric diseases such as Parkinson's and depression. In Parkinson's patients, sleep disorder is one of the main characteristics and one of the early and progressive symptoms, with approximately 75 percent of Parkinson's patients suffering from sleep disorders and insomnia. On the other hand, in patients with depression, sleep disorder is one of the main symptoms and can exacerbate the severity of the disease, with research showing that more than 90 percent of people with depression suffer from sleep problems. For this reason, extensive research is needed in this area to identify the cause of these diseases in the early stages and prevent secondary problems. Today, artificial intelligence and deep learning networks are widely used tools in medical science that can identify sleep disorders. Since visual and manual examination of physiological signals related to sleep disorders by specialists in hospitals and sleep clinics is time-consuming and exhausting, this thesis attempts to use an intelligent and highly accurate method to identify people with sleep disorders so that timely medical treatment can be provided by physicians. In addition to the mentioned advantages of this proposed method, one of the most important challenges in using LSTM neural networks for sleep disorder detection is the complexity of the structure and the high number of parameters of this neural network. LSTM networks, due to the complex nature of their internal architecture, have a significant volume of adjustable parameters that require considerable expertise and time to configure. For future research, in the first stage, novel approaches such as meta-heuristic algorithms like genetic algorithms, bee colony algorithms, etc., can be used for automatic tuning of the hyperparameters of this network, and in the next stage, more powerful and newer neural networks can be used to increase the accuracy of identifying healthy individuals and those with sleep problems.